\let\new=\newcommand
\new{\be}{\begin{equation}}
\new{\ee}{\end{equation}}
\new{\bfig}{\begin{figure}}
\new{\efig}{\end{figure}}
\new{\diff}{{\rm d}}
\new{\cE}{{\cal E}}
\new{\Jm}{{J^2_{max}}}
\new{\half}{\frac{1}{2}}
\new{\bp}{\mbox{\boldmath $p$}}
\new{\bx}{\mbox{\boldmath $x$}}
\new{\bv}{\mbox{\boldmath $v$}}
\documentstyle[referee,epsf,onecolumn]{mn}
\input psfig
\title[ Gravomagnetic Lensing by NUT Space ]
{Gravomagnetic Lensing by NUT Space}
\author[ Mohammad Nouri-Zonoz  and Donald Lynden-Bell] 
{M.~Nouri-Zonoz\thanks{Email:~nouri@ast.cam.ac.uk}, and 
	 Donald.~Lynden-Bell \thanks{Email:~dlb@ast.cam.ac.uk}\\ 
	 ~Institute of Astronomy, Madingley Road, Cambridge CB3 0HA}

\pagerange{\pageref{firstpage}--\pageref{lastpage}}
\begin{document}
\label{firstpage} 
\maketitle
\begin{abstract}  
Using the fact
that the null geodesics in NUT space lie on spatial cones, we
  consider
the gravomagnetic lens effect on light rays passing a NUT deflector. We show
 that this effect changes the observed shape, size and orientation of 
a  source. Compared to the Schwarzschild lens, there is an extra 
shear (a differential twist around the lens axis) due to the
 gravomagnetic field which shears the shape of the source. Gravomagnetic
 monopoles can thus be recognized by the spirality that they produce in 
the lensing pattern. 
\par All the 
results obtained in this case (magnification factor, orientation of
 images, multiplicity 
of  images, etc. )   
  depend on $Q$, the strength of the 
gravomagnetic monopole represented by NUT metric. One recovers the results
 of the usual  Schwarzschild lens effect  by putting
 this factor equal to zero.   
\end{abstract}
\section{INTRODUCTION}
The usual gravitational lens effect is based on the bending of light rays
passing a point mass $M$ (Schwarzschild lens) in  Schwarzschild spacetime .
The static  nature of the Schwarzschild metric implies that it can only
 produce gravoelectric
fields, but in  the case of NUT metric ( sometimes called the generalized
 Schwarzschild metric ), the existence of the cross term ``$d\phi dt$'' ,
 shows that 
 this space has also a  gravomagnetic field ${\bf B}_g$ (Demiansky $\&$ Newman 
1966).
\par A simple example illustrates what we will  
analyse relativistically in the following sections. Here we consider the 
deflection of an almost straight ray by the overall field of a stationary 
gravomagnetic monopole of strength $Q$ and mass $M$. There are two 
fields of the following form:

$${\bf E}_g=-{GM\hat {\bf r}\over r^2}\ \ \ \rm and \ \ \ {\bf B}_g=
-{Q\hat{\bf r}\over c r^2}$$
and the corresponding force per unit mass on the ray is 
$${\bf F}=-{GM\hat {\bf r}\over r^2}-{{Q{\bf v}\times \hat{\bf r}}\over cr^2}
,$$
where $v=c$ is the speed of the ray. The geometry is shown in Fig. 1.
The ray passes the lens parallel to x-axis with a large impact parameter $b$.
The normal gravitational force applies an inward radial force to the ray, but
there is another force due to the gravomagnetic field along the z-direction.

\begin{figure}
\centerline{\psfig{figure=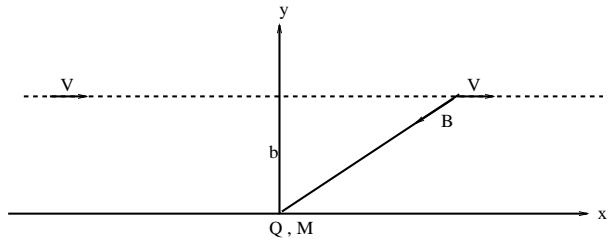,width=8cm,angle=-90}}
\caption{Light ray passing a gravomagnetic monopole at large impact 
parameter b.}
\end{figure}

The deflection due to the gravomagnetic force can be found by calculating the
 impulse applied to the ray, assuming that it is undeflected:
$$ \Delta {P_z}=\int \limits^{+\infty}_{-\infty}F_{z}dt=Qb\int \limits^
{+\infty}_{-\infty}
{dt\over{(b^{2}+{c^2}{t^2})}^{3\over 2}}={2Q\over cb}.\eqno (1)$$ 
The gravitational force deflects the ray in the x-y plane, but the 
gravomagnetic force deflects it out of the plane in the azimuthal direction
so  that the combined force twists the ray as the ray passes the monopole.

\section{NUT space and null geodesics}
The metric of the NUT space is given by (Kramer et al. 1980, Newman,Tamburino
$\&$ Unti 1963, Misner 1963) 

$${ds^2}=f(r){(dt-2l cos \theta d\phi)}^{2}-f(r)^{-1}{dr^2}-(r^2+l^2)(d\theta^2
+sin^2\theta d\phi^2),$$
where
$$f(r)=1-{2(mr+l^2)\over(r^2+l^2)},\eqno (2)$$
and $2l=Q$ is the strength of the gravomagnetic monopole represented by the
 NUT metric ($l$ is usually called the NUT factor or magnetic mass).
 Following the Landau and Lifshitz (1975) formulation of
stationary spacetimes and using the standard variational method (i.e.
$\delta\int ds^2=0$), the authors (1996) have shown that all the geodesics
 in NUT
 space, including the null ones, lie on cones defined through the following
 equation:
$${\bf L} + \varepsilon Q {\hat {\bf r}} = {\bf j} = \ {\rm const.}\footnote 
{We use the natural units in which $c=G=1$.} \eqno (3)$$

where ${\bf L}$ , ${\hat {\bf r}}$ and $\varepsilon$ are the angular momentum,
 unit radial vector and the energy constant of the orbiting particle and 
${\bf j}$
is a constant vector along the axis of the cone.
\begin{figure}
\centerline{\psfig{figure=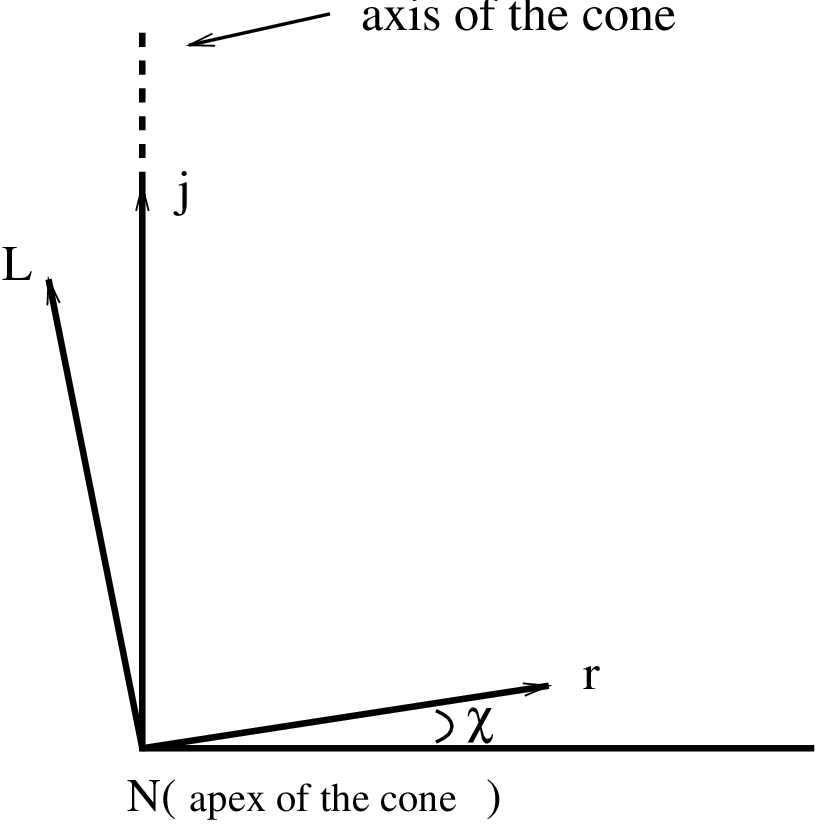,width=6cm,angle=-90}}
\caption{The relation between $\bf r$ , $\bf j$ and $\bf L$ }
\end{figure} 
Dotting equation (3) with  ${\hat {\bf r}}$ we find ${\bf j} \cdot {\hat {\bf
 r}} = \varepsilon Q$  showing that ${\hat {\bf r}}$ lies on a cone; similarly
 ${\bf L} \cdot {\bf j} = L^2 = {\bf j}^2 - \varepsilon ^2 Q^2 = \rm const.$,
 so that ${\bf L}$ also moves around a cone.
 Later we will be concerned with the semi-angle of the cone or its
 complement $\chi$. From Fig. 2 one can see that

$${\rm sin} \chi= {\varepsilon Q \over {|{\bf j}|}}={Q \over 
b(1+{Q^2\over b^2})^{1\over2}},$$
where $b={L\over \varepsilon}$ is the impact parameter. In the limit
$Q\ll b$, i.e. small $\chi$ we have
$$\chi\cong {Q\over b}.$$

\section {Bending of light rays in NUT space}

The equation of the  null geodesics in NUT space is  (Lynden-Bell $\&$ 
Nouri-Zonoz 1996)
$$f(r)^{-1}(\varepsilon ^2 -{\dot r}^2)-{L^2 \over {r^2}}=0 \ \ \ ,  \ \
 \dot r \equiv {dr \over {d \tau}}, \eqno (4)$$
where $\tau$ is an affine parameter.\\
Using the angle $\phi$ defined around the surface of the cone we have
$$ r^2 \dot \phi =L. \eqno (5)$$
Now using (5) we can integrate (4) and obtain
 
$$\Delta \phi=\int {L dr \over r^2(\varepsilon^2-{L^2 f(r)\over r^2})^
{1 \over 2}}.$$\\
We perform the above integral in the limit where $r^2 \gg l^2+m^2$, i.e. far
away from the event horizon (which is the case for the observed gravitational
 lenses) and where we have 

$$\Delta \phi=\int\limits^{r_{min}}_{\infty} {L dr\over r^2[\varepsilon ^2 -
{L^2 \over r^2}(1-{2m \over r})]^{1\over 2}}.$$
Changing to the variable $u={1\over r}$ we have 
$$\Delta \phi=-\int\limits^{u_{max}}_0 {du\over (\varepsilon ^2 -
L^2 u^2+2mL^2 u^3)^{1\over 2}}.$$ 
Calculating this integral to the first order in $mu$
one can see that the deflection angle on the cone is

$$\alpha=2\Delta \phi - \pi={4m\varepsilon \over L} = {4m\over b}.\eqno (6)$$
which is the same as in the Schwarzschild case but with the difference that 
now $b={L \over \varepsilon}$  is the impact parameter defined on the cone.

\section {Geometry of Lensing in NUT space}

The gravomagnetic lensing configuration is shown in Figs 3-5. A NUT lens
 is located at distances $D_d$ and $D_{ds}$ from an observer O and a source
S respectively. The plane of the sky (or the lens plane) in which all the
 observations are made is the plane normal to $D_d$ at N.

\begin{figure*}
\centerline{\psfig{figure=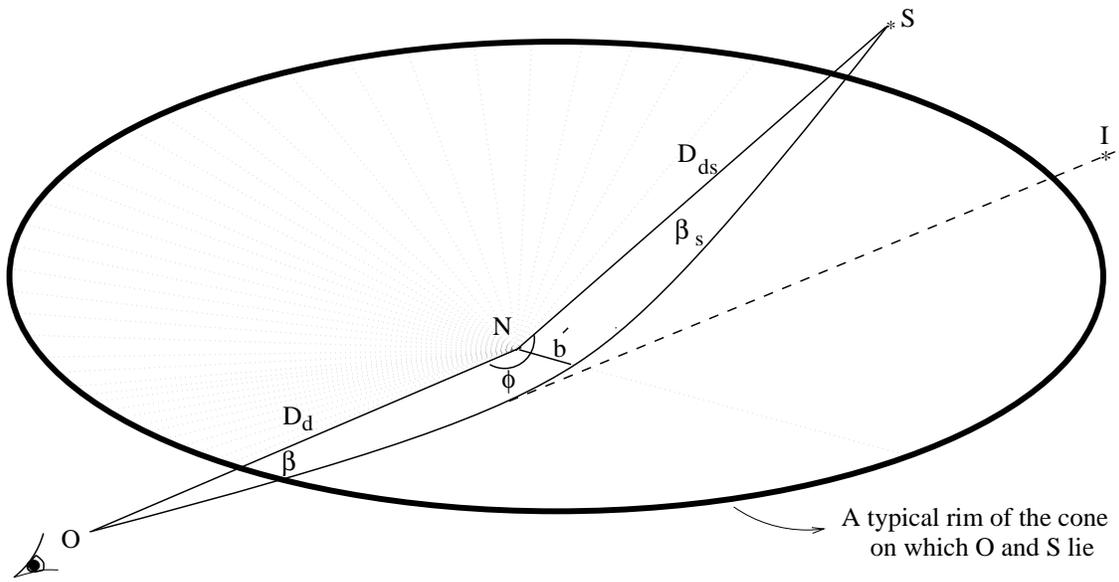,width=.9\textwidth,angle=-90}}
\caption{ A light ray (curved line) from $S$ propagating in NUT space on a cone
 centered at $N$ and passes the lens at distance $b^\prime$, where 
$b^\prime \gg event\; horizon$ (To the 
first order in $\chi$, $b^\prime=b$, where $b$ is the impact parameter). Note
that S and O lie on the cone but not I, because OI is the straight light
 ray that the observer traces back to the image . }
\end{figure*}

For small deflection angle  $\alpha$,  $b \ll {\bf D}_d$ and ${\bf D}_{ds}$,
 so from figure 3 we have
$$\beta \cong {b \over D_d } \ \ \ , \ \ \ {\beta_s \cong {b\over D_{ds}}}. 
\eqno(7)$$
Using these two angles one can see that

$$ \phi=\pi+\alpha-\beta-\beta_s = \pi + {4m\over b}-{b\over D_d}(1+{D_d \over 
D_{ds}}), \eqno(8a)$$and 

$$\eta={\phi \over \rm cos \chi}={[\pi + \alpha - {b\over D_d}(1+{D_d \over 
D_{ds}})] \over \rm cos \chi}, \eqno(8b)$$
where $\phi$ is the angle between generators $D_d$ and $D_{ds}$ around the
 surface of the cone  and $\eta $ is the  projection  of $\phi$ onto the plane
 normal to the axis of the cone at its apex.
\section {Lens equation}
The lens equation is basically the mapping of the source position to 
the image position on the lens plane. To find this mapping in the NUT case
, we first need to find
 the angle between 
position vectors of the source and image (${\bf r}$ and ${\bf r}^{\prime}$)
 on the lens plane (see Fig. 5).
\par As can easily be seen from the geometry in Fig. 5, the positions of
 the source and image on the lens plane are the intersections of
this plane with lines OS and OI respectively (Points F and G). One should
 note that  the point G is not on the cone but on the plane normal to the axis 
 of the cone at its apex.\\
From Fig. 5 one can see that the angle between  the position vectors of
 the source and image is the same as the angle between vectors
 ${\bf r}^\prime$ and 
$\bf {NM}$. Where $\bf {NM}$ is the component of $\bf D_{ds}$ on the lens 
plane. Therefore, representing the vector $\bf {NM}$ by $\bf R$, we have 
 
$${\bf r}^{\prime} . {\bf R} ={\bf r}^{\prime} . [{\bf D}_{ds} - ({\bf D}_{ds} . 
{\hat{\bf D}}_d){\hat{\bf D}}_d]={\bf r}^{\prime} . {\bf D}_{ds} .\eqno (9a) $$
\\
On the other hand

$${\bf r}^{\prime}= r^{\prime}({{\bf j}\times {\bf D}_d \over j D_d \rm cos 
\chi}).\eqno (9b)$$

$$R=D_{ds}  \rm sin \theta .\eqno (9c) $$ 
where $\theta $ is the angle between generators $D_d$ and $D_{ds}$ in the 
triangle ONS. Substituting equations (9b,c) in (9a) and using a spherical
 coordinate  system centered at N one can show that

$$D_{ds}\rm sin\theta \rm cos\psi = {{({\bf D}_d \times {\bf D}_{ds}) . 
{\bf j}}\over {j D_d \rm cos \chi}}.$$ But

$${({\bf D}_d \times {\bf D}_{ds}) . {\bf j}}=|j|D_d D_{ds} \rm cos^2\chi 
\rm sin\eta .$$
Therefore
$$\rm cos\psi={\rm {cos^2 \chi\rm sin \eta}\over \rm sin \theta}. \eqno (10)$$
Using the same spherical coordinates and equations (8a,b) one can show that

$$ \rm cos\theta =\rm cos^2 \chi \rm cos\{ {(\pi+\alpha) -\beta [1+(D_d/ D_{ds})]\over \rm cos\chi}\}+ \rm sin^2 \chi .\eqno(11)$$

Equations (10) and (11) together determine $\psi$ the angle between the
 position vectors of the source and image in the lens plane. In what follows 
we will be concerned with the small angle (i.e. $\alpha, \beta, \chi \ll 1$)
approximation to the expressions (10) and (11) which are 
$$ \rm cos\theta \cong -1+2\chi ^2 +{1\over 2}[\alpha - 
\beta (1+{D_d\over D_{ds}})],\eqno(12a)$$ and 

$$ \rm cos\psi \cong {\beta (1+{D_d \over D_{ds}})- \alpha -{{\pi \chi^2} 
\over 2}\over{[4\chi^2 +{(\alpha-\beta(1+{D_d\over D_{ds}}))}^2]}^{1\over 2}}. 
\eqno(12b)$$
\begin{figure*}
\centerline{\psfig{figure=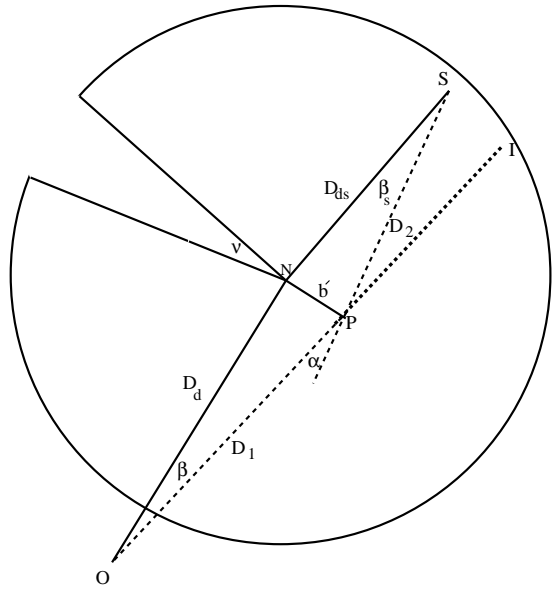,width=.45\textwidth,angle=-90 }}
\caption{ Open, flattened cone and the light ray (dashed line) which is 
deflected at $P$ passing the NUT lens. $\nu=2\pi(1-{L\over 
(L^2+\varepsilon ^2 Q^2)^{1\over 2}})$ and $\alpha$ are the deficit 
and bending angles respectively.}
\end{figure*}
\begin{figure*}
\centerline{\psfig{figure=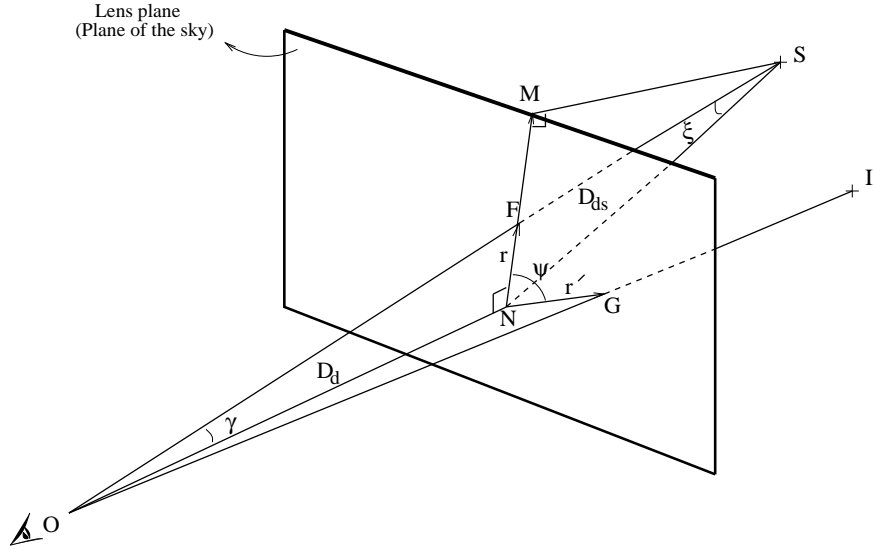,width=.7\textwidth,angle=-90 }}
\caption{ Lens plane and the positions of source S, image I and the observer
O. }
\end{figure*}
The following three limiting cases are worth considering:\\
a) $\chi=0$ for which we get $\psi=0$ for $\alpha > \beta$ and $\psi=\pi$ for 
$\alpha < \beta$,\\
b) $\alpha=\beta =0$ and $\chi\ll 1$ for which we find that $\psi
={\pi \over 2}$,\\
and finally,\\
c) $\beta \gg \alpha\; \& \;\chi$ from which we have 
$\psi={2 / {\bar\beta}}
({q/ b})$ where  $\bar \beta=\beta (1+{D_d\over D_{ds}})$.\\
The above three results are all in agreement with what one expects from simple
geometrical considerations.\\
Having found the angle $\psi$, the lens equation which maps the position 
of the source to that of the image, can be written as follows:
$$ {\bf r}=\kappa A {\bf r}^\prime\;\;\;\;\;\;\;\;\;\;\;\;\;\;\; \kappa=
{r\over r^\prime}$$ or
$$\left\{ \matrix{x=\kappa(x^\prime\rm cos\psi - y^\prime \rm sin\psi)\cr
y=\kappa(x^\prime\rm sin\psi + y^\prime \rm cos\psi).\cr}\right.\eqno(13)$$
where $\kappa$ is given in equation (17) and ${\bf r}$ and ${\bf r}^\prime$ 
denote the position vectors of  the source 
and image in the lens plane with the coordinates (x,y)(see Fig. 5).
\section {The magnification factor}
To see what happens to a bundle of rays in passing a NUT lens, we consider 
the magnification of an infinitesimal circular source centered at 
 ${\bf r}$; i.e. (Fig. 6)\\
$${(X-x)}^2+{(Y-y)}^2 ={\delta x}^2+{\delta y}^2 ={s^2}\;\;\;\;\;\;\;\;\;\;\;
 s\ll 1 .\eqno(14)$$\\

Now by varying equations (13) and using (14) we have 
\begin{eqnarray}
{\delta x}^2+{\delta y}^2 ={s^2}=
(\delta\kappa)^2({x^\prime}^2+{y^\prime}^2)
+\kappa^2({\delta x^\prime}^2+{\delta y^\prime}^2)+ \kappa^2 
{(\delta \psi)}^2{r^\prime}^2 +\;\;\;\;\;\;\;\;\;\;\;\;\;\;\;\;\;\;\;
\; \nonumber \\
\;\;\;\;\;\;\;\;\;\;\;\;\;\;\;\;\;\;\;\;\;\;\;\;\;\;\;\;\;\;\;\;\;
\;\;\;\;\;\;\;\;\;\;\;\; +2\kappa(\delta \kappa)({x^\prime} 
\delta x^\prime +{y^\prime}
\delta y^\prime)+2\kappa^2 ({x^\prime} \delta y^\prime -{y^\prime}
\delta x^\prime)\delta\psi \;\;\;\;\;\;\;\;\; (15) \nonumber
\end{eqnarray}
On the other hand, using the geometry of lensing as shown in Fig. 5
\begin{figure*}
\centerline{\psfig{figure=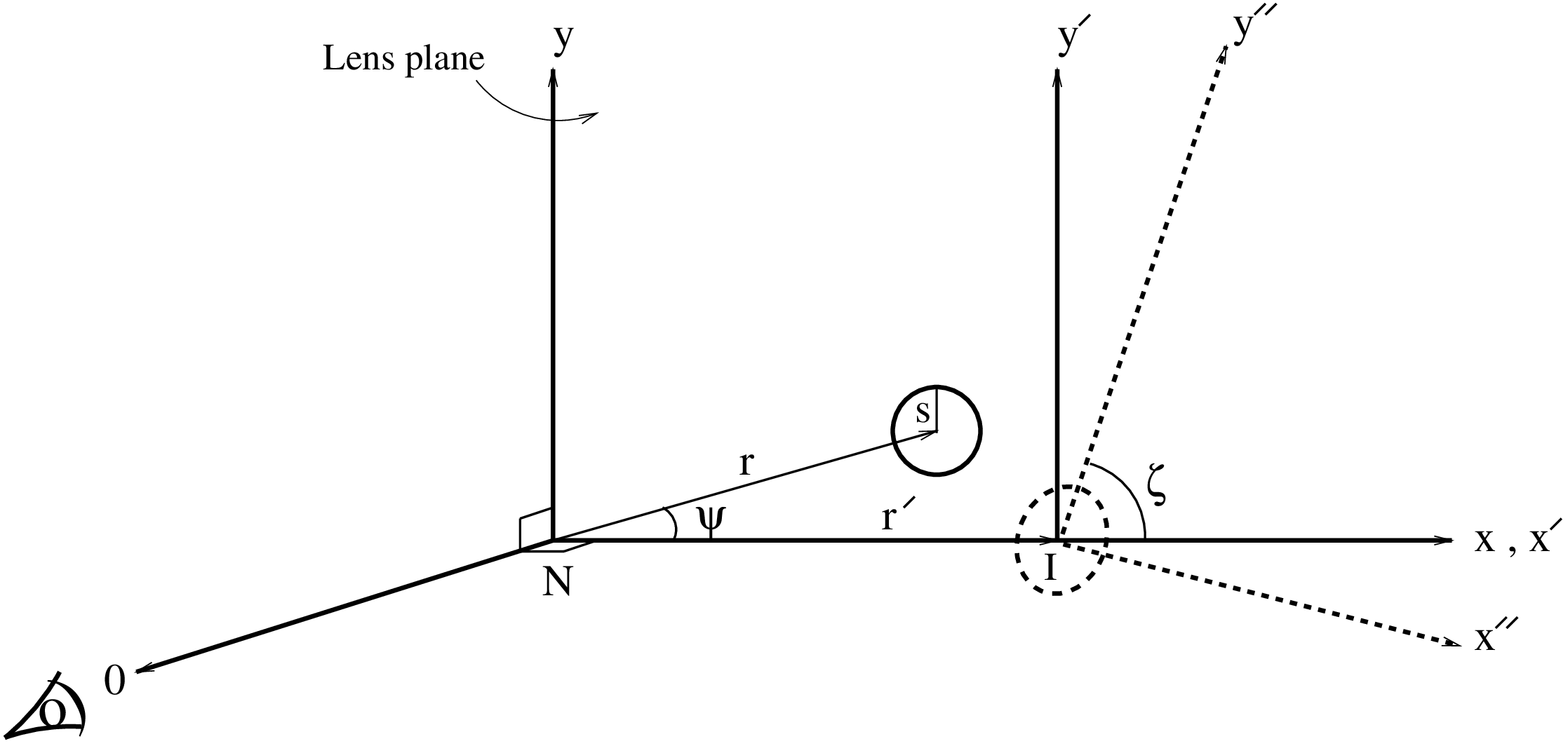,width=14cm,angle=-90 }}
\caption{Small circular source at $\bf r$ lensed to an ellipse at $\bf
 r^\prime$  }
\end{figure*} 
for small angles, we have\\
$$r^\prime = \beta D_d \cong b,\eqno(16a)$$and

$$r=\gamma D_d =D_d{{\{4\chi^2 +{[\alpha-\beta(1+{D_d\over D_{ds}})}]^2\}}^
{1\over 2}\over (1+{D_d \over D_{ds}})}.\eqno(16b)$$
where in finding the angle $\gamma$ we used equation (12a) and the fact that
(fig.5)
$$ \gamma D_d=\xi D_{ds}\;\;\;\;\;\;\;\;\;\; , \;\;\;\;\;\;\;\;\;\;\;\;\ 
\gamma + \xi=\pi-\theta .$$
So
$$\kappa={r\over r^\prime}={\gamma \over \beta}={{{[4\chi^2 +{(\alpha-\bar 
\beta)}^2]}^
{1\over 2}}\over {\bar \beta}},\eqno(17)$$ 
and
$$\delta \kappa=({\delta b\over b\bar \beta}){(2\alpha {\bar \beta}-8 
\kappa^2 -2 \alpha^2) \over {[4\chi^2 +{(\alpha-\bar \beta)}^2]}^{1\over 2}}
,\eqno(18)$$
where
$$\bar \beta \equiv \beta (1+{D_d\over D_{ds}}),$$
$$\delta b \cong \delta r^\prime={(x^\prime \delta x^\prime + y^\prime
 \delta y^\prime)\over {r^\prime}},\eqno(19)$$
and we have used the fact that for small angles $\alpha$,$\beta$ and $\chi$ 
$$\delta \beta=\beta({\delta b\over b})\;\;\;\; , \;\;\;\; \delta \alpha = 
-\alpha ({\delta b \over b})\;\;\;\; , \;\;\;\; \delta \chi =-\chi
 ({\delta b \over b}).$$
Using equations (19) and (12) one can show after some manipulation that
$$\delta \psi = - {4\beta \chi \over {[4\chi^2 +{(\alpha-\bar \beta)}^2]}}
({\delta b \over b}). \eqno(20)$$
Substituting (17), (18), (19) and (20) in (15)  and after a simple but long
calculation one finds that to the lowest order in angles and increments 
$$\displaylines{s^2={4 \alpha(x^\prime \delta x^\prime + 
y^\prime \delta y^\prime)^2 \over b^2 \bar\beta} + 
{[4\chi^2 +{(\alpha-\bar \beta)}^2] \over {\bar \beta}^2}
[(\delta x^\prime)^2+(\delta y^\prime)^2]+\cr \hfill{}+ 
{8 \chi \over b^2 \bar \beta}(x^\prime \delta y^\prime - 
y^\prime \delta x^\prime)(x^\prime \delta x^\prime + 
y^\prime \delta y^\prime),\hfill{(21)}\cr}$$
Putting $y^\prime =0$, without loss of generality, the equation of the image
 becomes 
$$(A+B){X^\prime}^2 + B {Y^\prime}^2+2CX^\prime Y^\prime =1, \eqno(22)$$
where
$$A={4\alpha \over {\bar\beta} s^2}\;\;\;\;\;\;\; , \;\;\;\;\;\;\;
 C={4\chi \over {\bar\beta} s^2},$$
$$\eqno(23)$$
$$B= {{[4\chi^2 +{(\alpha-\bar \beta)}^2]} \over {s^2 \bar \beta}^2},$$
and we put $\delta x^\prime=X^\prime$ and $\delta y^\prime=Y^\prime$.\\
One can see that (22) is the equation of an ellipse in $X^\prime Y^\prime$
coordinates. By definition the magnification '$\mu$' is the ratio of the area
of the image to the area of the unlensed source, so
$$\mu= {the\; area\; of\; the\; image \over the\; area\; of\; 
the\; sourse}={1\over ({\lambda_+}{\lambda_-})^{1\over2}}
={1\over [(1-{\alpha^2\over \bar \beta ^2})
(1-{\alpha^2\over \bar \beta ^2}-{8\chi^2 \over \bar \beta ^2})]^{1\over 2}
},\eqno(24)$$
where $\lambda$s are the eigenvalues of the rotation matrix that transforms
 the ellipse into its principal axes (${x}^{\prime\prime}{y}^
{\prime\prime}$).(Fig. 6)\\
As it can easily be seen for $\chi=0$ one recovers the known result of the
 Schwarzschild lens.
The orientation of the ellipse is determined by the angle $\zeta$ shown
in figure 6
$$tan\zeta = {2\chi \over {(\alpha ^2 +4{\chi}^2)}^{1\over 2}- \alpha },$$
or in terms of the strength of the gravomagnetic monopole $Q$ 
$$tan\zeta = {Q \over {(4m^2+Q^2)}^{1 \over 2}-2m}.$$
For $Q=0$, as expected, we recover the result of the Schwarzschild lens
 which is $\zeta={\pi \over 2}$. So the presence of the NUT lens, not only
 changes the shape and size of the source but also its orientation and it 
is this last change which produces the spirality effect in the lensing 
pattern. For example for $Q_1={m \over 2}$ and $Q_2=m$ we get  
$\zeta_1=82.98$ and
$\zeta_2=76.71$. Therefore the effect of the gravomagnetic field even in these
 extreme cases is quite small. This twist in the observed image would not be 
an unexpected effect if one looks at the null 
tetrad formulation of null geodesics in NUT space (see the appendix ).
\section {Multiplicity of Images}
One can easily see that the limit $\chi=0$ of the lens equation recovers
the results of the Schwarzschild lens concerning the number of the images.
In this section we will consider the number of the images for a given
 position of the source when $\chi \ne 0$.\\
Writing equation (17) and using the fact that to the lowest order in $\chi$,
  $r^\prime =b $, we have 
$${r\over b }={[{4Q^2 \over b^2}+({4m\over b}-bA)^2]^{1\over2}\over bA},$$
where $A={D_{ds} + D_d \over D_{ds}D_d}$. Using the above equation one 
arrives at
the following equation for $b$ 
$$A^2 {b^\prime}^2 + B b^\prime + 4C =0 \;\;\;\;\;\; , \;\;\;\;\;\;
 b^\prime=b^2,\eqno(25)$$
where 
$$B=-(8mA + r^2A^2),$$
$$C= 4Q^2 + 16m^2.$$
Solutions of (25) are 
$$ b=({4m\over A} + {r^2 \over 2}\pm {1\over 2A^2} \sqrt \Delta )
^ {1 \over 2}, \eqno(26)$$
where
$$\Delta = r^4A^4 + 16mA^3r^2- 16A^2Q^2. \eqno (27)$$
We see that, in general, for a given $r$ there can be two solutions for b
(when $\Delta > 0$ or equivalently $r^2 > {r_0}^2$; see below).
When $\Delta=0$, there is one solution for b which happens at  
$$r^2={r_0}^2={8m\over A}(1+{Q^2 \over 4m^2})^{1\over 2}- 
{8m \over A}.\eqno(28)$$
In this case using (27) and (28) one can see 
that for $Q=0 \rightarrow r=0$ and therefore 
$$b=({4m\over A})^{1\over 2}.$$ This is the solution which corresponds to the
Einstein rings in the Schwarzschild lensing. In figures 7 and 8  a 
horizontal line of point sources and its images for the 
two cases $\alpha <\beta$ and $\alpha >\beta$ are shown and 
Figs 9 - 13 show randomly distributed small circular sources ( for both
$m=Q=1$ and $m=0$
, $Q=1$ cases with $D_{ds} \rightarrow \infty$ ) in the plane 
of the sky and their lensing pattern. One should note that when $m=0$, 
according to equatin (6), $\alpha =0$ to the first order in $m\over r$ but by
performing integral (5)  
to the lowest order in $l\over b$ 
(for $m=0$) we can show that $\alpha ={\pi\over 2}{l^2\over b^2}$ 
 (Appendix B). To have a better idea of lensing, some
of the point sources as well as the circular ones and their corresponding 
images are labled. For $\alpha <\beta$
all the circles are magnified but for $\alpha >\beta$ 
( Figs 11 and 13 ) all
the circles are reduced. The spirality of the lensing pattern can be seen
 more clearly near to the center of the deflection. What the observer will 
actually see is the superposition of the two cases 
$\alpha <\beta$ and $\alpha >\beta$ ( Figs 10 and 12 ).\\
As is clear from equation (26), for $\Delta < 0$ (or equivalently
$r^2 < {r_0}^2$ ) there are no solutions for $b$, in other words an observer
at a given position will not be able to see the  source, if
it lies inside a circle of radius less than $r_0$. (point source $a$ of
 figure 7  and circular source $b$ of Fig. 9 are  absent in the lensing
 pattern of  the Figs 8 and 12 because of this and 
the same reason applies to  the incomplete images in Figs 10 - 13 ). This 
does not mean that there is a gap or a hole (in the case of a uniform
distribution of point sources) in the  observed image, because
 as  is shown in Figs 10 and 12, the reduced images ( along with the 
images of  the sources nearer to the lens which have smaller $r_0$ by 
equation (28) ) will fill in the expected hole area.

\subsection{Part of the sky which can not be seen by a specified observer}
As we mentioned in previous section equation (25) has no solutions for 
$r < {r_0}$. This means that for an observer with a given distance $D_d$
from the lens a part of sky is unseeable. Expanding equation 
(28) in powers of $Q^2\over m^2$ and keeping the first order term we have
$${r_0}^2={Q^2\over Am}$$
Figures 14 and 15  show $r_0$ as a function of $D_{ds}$, 
the distance between the source and the lens.

\subsection{Critical orbits and the dark area of NUT black hole}
To find the area at the NUT lens where due to the trapped null geodesics the 
observer sees (looking directly at NUT hole) a small black sphere we need 
to find the critical orbits which 
are specified by
$${{\rm d}r\over \rm d\phi}=0 \;\;\;\;\;\;\;\;\;\;\;\;\ {{\rm d}^2r
\over {\rm d }\phi^2}=0\eqno(29)$$
Using (5) one can see that the above two equations
are equivalent to the following two equations 

$$G(r)={1\over r^2+l^2}({dr\over d\phi})^2=0 \equiv (r^2+l^2)^2-
(r^2-l^2-2mr)b^2=0\eqno(30)$$
$${d^2 r\over d\phi^2}=0 \equiv G'(r)= r^4-l^4-l^2 b^2-mrb^2=0\eqno(31)$$
Now substituting for $r^4$ in (31) from (30) we have
$$r^2(2l^2-b^2) + 3mb^2 r+2l^2(b^2+l^2)=0 $$
Solving this for $r$ we find
$$r={3mb^2+\sqrt{9m^2 b^4 + 8l^2(b^2+l^2)(b^2-2l^2)}\over 
2(b^2-2l^2)}\eqno(32)$$
Now to find the impact parameter which is basically the radius of the dark 
area we need $r$ in the two limiting cases in which a) $l\ll m$ and 
b) $m\ll l$.\\
For $l\ll m$ we find from (32)
$$r=3m(1+{2\over 9} {l^2 \over m^2})\eqno(33a)$$
substituting this into equation (31) and solving for $b^2$ we have
  
$$b^2=27m^2[1+{1\over 3}({l^2 \over m^2})]\eqno(33b)$$
It is clear that these equations give the right answer for the 
Schwarzschild case i.e. for $l=0$
$$ r=3m \;\;\;\;\;\;\;\;\;\;\; , \;\;\;\;\;\;\;\;\;\;\; b=3 \sqrt{3}m 
\eqno(34)$$
but because we took the limit in which $l\ll m$, the above results do not 
give the exact value for the pure NUT case (m=0), for this we need to 
repeat the above procedure for the $m\ll l$ limit where from (32) we have 

$$r=\sqrt{3}\; l\; [1+{m\over l} ({6\over 27} D+{2 \over \sqrt{3}} -
{2\over 3})]\eqno(35a)$$
$$b=2\sqrt{2}\; l\; [1+D(m/l)]\eqno(35b)$$
 in which $D={5\sqrt{2}+48 \over 24 \sqrt{3}}$.
These equations are in agreement with the results of 
the pure NUT case which are
$$ r=\sqrt{3}\;l \;\;\;\;\;\;\;\;\;\;\; , \;\;\;\;\;\;\;\;\;\;\; 
b= 2\sqrt{2}\;l\eqno(36)$$ 
An interpolation formula between equations (33a,b) and (35a,b) which are
good when either {\it m} or {\it l} is small is given by

$$r^2_{_N}={m^2(9m^2+13 l^2)+ 
3l^3 \; [l+2m({6\over 27} D+{2\over \sqrt{3}}-{2\over 3})]\over l^2+m^2}
\eqno(37a)$$
$$b^2_{_N}={9m^2(3m^2+4l^2)+8l^3(l+Dm)\over (m^2+l^2)}\eqno(37b)$$
Equations (37) show that a circular orbit of radius $r_{_N}$ is an unstable 
null geodesic and null geodesics arriving from infinity with an impact 
parameter less than $b_{_N}$ cross the event horizon and get trapped. 
Therefore $b_N$ is the radius of the 
small black sphere 
the observer sees when he looks directly at the NUT black hole. 
\section {Time delay between images}
As we saw in the last section  gravomagnetic lensing by NUT space can 
produce multiple images of a source. Different images correspond to different
impact parameters and path lengths and therefore to different travel times.
\par As in the usual lensing two different effects are responsible for this
delay, namely  the difference in path lengths and the difference in the 
gravitational potential traveresed by different light rays. In what follows 
we calculate the time delay between the arrival time for an image in the
 presence of the NUT lens and in its absence, taking into account both of 
the above contributions.
\subsection { Geometric time delay }
Geometric time delay by definition is 
$$ (\Delta t)_{geo.} = \Delta l / c. \eqno(38)$$
where $\Delta l$ is the extra path length of the deflected light ray relative
to an unperturbed null geodesic in the absence of the lens. Using Figs 3.3-3.5
 and the geometrical relations defined there we see that 
$$\Delta l=(D_1+D_2)-|{\bf D}_{ds} -{\bf D}_d|,$$
where
$$D_1={cos({\alpha \over 2}-\beta) \over cos{\alpha \over 2}}D_d=(1+{\alpha^2
\over 8})[1-{1\over 2}({\alpha \over 2}-\beta)^2]D_d.$$
   
$$D_2={cos({\alpha \over 2}-\beta_s) \over cos{\alpha \over 2}}D_{ds}=(1+{\alpha^2\over 8})[1-{1\over 2}({\alpha \over 2}-\beta_s)^2]D_{ds}, $$ 
and

$$|{\bf D}_{ds} -{\bf D}_d|=(D_{ds}+D_d)-{D_d D_{ds} \over D_d+D_{ds}}[2\chi^2+{1\over 2}(\alpha - \bar\beta)^2)],$$ is the norm of the vector 
joining the source and the observer.\\
Therefore 
$$\Delta l ={1\over 2}D_{ds} (\beta_s \alpha -{\beta_s}^2)+
{1\over 2}D_{d} (\beta \alpha -{\beta}^2)+
{D_dD_{ds} \over D_d+D_{ds}}[2\chi^2+{1\over 2}(\alpha - \bar\beta)^2)].$$
Substituting for $\beta$ and $\beta_s$ from equation (7) we have
$$ (\Delta t)_{geo.} = \Delta l / c ={1\over c}[{1\over 2}{D_d D_{ds}\over 
D_{ds}+D_d}\alpha^2+2({D_dD_{ds}\over D_d+D_{ds}})\chi^2].\eqno (39)$$
\subsection {Potential delay (Shapiro's delay)}
As we showed in section 1.2.5, potential delay in NUT space can be 
calculated by using the Fermat generalized
principle for stationary spacetimes (Landau $\&$ Lifishitz 1975, Schneider, 
Ehlers $\&$ Falco 1992) which reads;
$$\delta t=\delta \biggl( {1\over c} \int {dl\over[{g_{00}}^{-1}+(A_\alpha 
{dx^\alpha \over dl})]^{-1}}\biggl)= 0\;\;\;\;\;\;\; \alpha=1,2,3.  \eqno(40)$$
where $A_\alpha={-g_{0\alpha}\over g{00}}$ and  $dl$ is the Euclidean distance 
element along the light trajectory.\footnote{Note that the calculations are 
done in the limit where
 $r^2\gg Q^2,m^2$ i.e far away from the horizon.}\\
Equation (49) shows that ${{g_{00}}}^{-1}+A_\alpha {dx^\alpha \over dl}$ can be
thought of as the refractive index, $n$, of the spacetime under consideration.
(in this case the NUT metric). Using this fact, the potential delay will
 have the following form
$$(\Delta t)_{pot.}=\int {dl \over {c\over n}}- \int {dl\over c},$$
or for the NUT space 
$$(\Delta t)_{pot.}={1 \over c}\int^{Observer}_{Source} 
[({1-{2m\over r}})^{-1} -Q\rm cos \theta {d\eta \over dl}]dl - 
{1\over c}\int^{Observer}_{Source}dl$$
where $r$ , $\theta$ and $\eta$ are the usual spherical coordinates. 
Using the facts that $m\ll r$, $\rm cos\theta =\rm sin\chi$ and $\eta= 
 {\phi \over\rm cos\chi}$ we have
$$(\Delta t)_{pot.}={1\over c}\int^{Observer}_{Source}{2m\over r}dl + {\chi^2
\over c}(\pi+\alpha -\bar \beta)b, \eqno(41)$$
where we used equation (8a) and the fact that $\rm tan\chi \cong\chi
={Q\over b}$.
 It can easily be shown that the first term to the first order in $m\over r$ 
is the potential delay for Schwarzschild metric, i.e.
$${1\over c}\int^{Observer}_{Source}{2m\over r}dl={2m\over c}{\it ln} 
{4D_dD_{ds}\over b^2}.\eqno(42)$$
Therefore the total time dealy in  NUT lensing is 
$$\displaylines{(\Delta t)_{tot.}=(\Delta t)_{geo.}+(\Delta t)_{pot.}
 ={1\over c}[ {D_dD_{ds}\over 2D_{ds}}+2({D_dD_{ds}\over D_d+D_{ds}})^
2\chi]^2 + \cr \hfill{} +{2M\over C^3}{\it ln}{4D_dD_{ds}\over b^2}+
{\chi^2\over c}(\pi+\alpha - \bar\beta)b \hfill{(43)}\cr }$$
Again we note that (43) reduces to the Schwarzschild lensing time delay 
when $\chi=0$.\\
Note that if the calculation were to be done in R-W spacetimes the D's would 
have been
 distances determined by the apparent size i.e. $D_d$ is evaluated at the 
time when light passes the lens ( redshift $Z_d$ ) and $D_{ds}$ is evaluated 
when light is emitted. (redshift $Z_{ds}$)
\section {Discussion}
We have studied the lensing of light rays passing a NUT deflector in the
context of gravoelectromagnetism. We have shown that there is an extra shear
 due to the presence of the gravomagnetic field of the NUT space which shears 
the shape of the source.
It was shown that this effect produces a spirality in the lensing pattern of 
a random distribution of circular sources  located at infinity. This 
spirality effect about a NUT lens is very characteristic and is not 
displayed in normal lensing and the gravomagnetic lens due to a rotating 
object seen pole on does not show it because the twist of the ray as it 
approaches such a lens is cancelled by the opposite twist as it recedes.
Thus  the discovery of a spiral shear field about a lens would 
indicate, at least in principle, the presence of a grvomagnetic lens.
\par The observability of gravomagnetic lenses is also discussed in the 
context of the constraints on magnetic masses (Mueller $\&$ Perry 1986). 
It was shown by Misner (Misner 1963) that one can
remove the string singularity in NUT space by using two different coordinate 
patches to cover the northern $(0\leq \theta \leq {\pi\over 2})$ and southern 
$({\pi\over 2}\leq \theta \leq \pi)$ hemispheres at the cost of identifying the
time coordinate with period $8\pi l$.\footnote {It was shown that this 
periodicity condition is a natural consequence of quantum mechanical 
considerations and  holds for any spacetime that has magnetic mass
 (Dowker 1974, Hawking 1979).} One can show that this 
periodicity is equivalent to the existence of closed timelike curves 
which in turn violates the causality. Mueller and Perry then discuss 
that since the Universe is not observed to be periodic on the scales less 
than $10^{10}$ yr, any magnetic mass is presumably very large and at least 
$10^{60}$ Planck masses. This means that the characteristic scale size 
for such a mass would 
be the Hubble scale or larger. This large value for magnetic masses makes the 
lensing effect the most promising way to look for them. Although the 
expectation must be small, such effects should be looked for by those studying
gravitational lenses.
\section*{ACKNOWLEDGEMENTS}
The authors would like to thank H. Ardavan  for useful 
discussions and comments. MNZ thanks the Ministry of culture and Higher 
Education of Iran for financial support and is grateful to L. W. Chen 
and C. Rola for helping him with computer graphics.
\section*{Appendix A: Null tetrads and the geometry of null geodesics}
Consider the family of null geodesics of a metric $g^{ab}$. Let $l^a={dx^a
\over du}$ denote the null tangent vectors to these geodesics, where $u$ is 
an affine parameter along the geodesic. At each point one can define a 
complex, quasi-orthonormal null tetrad by the equation (Sachs 1961) 
$$g^{ab}=2l^{(a}n^{b)}- 2{\bar m}^{(a}m^{b)},$$
where $n^a$ is another real null vector and $\bar m^a$ is the complex 
conjugate of $m^a$. The above equations holds if and only if the 
quasi-orthonormality relations 
$$l^a n_a=-{\bar m}^a m_a=1,$$
$$ l^al_a=n^an_a=m^am_a=l^am_a=n^am_a=0,$$
hold. The vectors $(l^a, n^a, m^a, {\bar m}^a)$ define what is called a $ 
null\;\; tetrad$.
\par  Using this null tetrad one can find some of the geometrical
 properties of the null geodesics, e.g. define the following two complex
 quantities (Hawking 1973) 
$$ \rho=l_{a;b}m^a{\bar m}^b,$$
$$ \sigma=l_{a;b}m^am^b.$$
The real part of $\rho$ measures the average rate of convergence of nearby 
null geodesics in space $g^{ab}$ and therefore  is responsible for 
the magnification or 
reduction  of the observed image. The imaginary part of $\rho$  measures 
the twist or rate of rotation of neighbouring null geodesics i.e. this is
 the part
responsible for the spirality in the lensing pattern shown in the text.
One can easily calculate $\rho$ for the Schwarzschild and NUT metrics: 
$$\rho_{NUT}={-r+{\it i}l\over {r^2+l^2}}$$
$$\rho_{Schw.}=-{1 \over r}$$
We note that the same correspondence between the two metrics also applies here 
i.e.
$$\rho_{Schw.}=\lim_{l\to 0} \rho_{NUT}.$$
The above relations show that in NUT case not only the size but also the 
orientation of the observed image changes ( ${\rm Im}\rho={l\over r^2+l^2}$)
but in the Schwarzschild case there is no twist in the observed image because
${\rm Im}\rho=0$. 
The quantity $\sigma$ measures the rate of distortion or shear 
of the null geodesics. The effect of shear makes an originally small circular 
 cross section of a bundle of rays become elliptical. 
\section*{Appendix B: Bending angle  for pure NUT ($m=0$)}
In this appendix we  calculate the bending angle for a pure NUT case 
i.e. for m=0. Writing integral (5) for pure NUT we have
$$\Delta \phi=\int {L dr \over r^2\left( \varepsilon^2-{L^2 \over r^2}
(1-{2l^2 \over r^2}) \right) ^{1 \over 2}}.$$\\
where $(1-{2l^2 \over r^2})=e^{-2\lambda}|_{m=0}$. Now expanding the 
integrand in
powers of $l^2 \over r^2$ and keeping the lowest order term we have
$$\Delta \phi=b \int {dr \over (r^4-b^2 r^2+2l^2b^2)^{1 \over 2}}$$
where $b=L / \varepsilon$ is the impact parameter.
One can write the above integral in the following form
$$\Delta \phi=b \int^{r_{min}=b}_\infty {dr \over 
\left( [r^2-a^2][r^2-d^2] \right) ^{1 \over 2}}\eqno(B1)$$
where
$$a^2={2l^2 b^2 \over b^2/2 - b/2 \sqrt{b^2-8l^2}}\cong b^2\eqno(B2)$$
$$d^2=b^2/2 - b/2 \sqrt{b^2-8l^2}\cong 2l^2\eqno(B3)$$
Now due to the fact that the condition $r_{min} \geq a > d > 0$ 
is satisfied the above 
integral is an elliptic one of the first kind i.e. (equation 12 page 
246 of Gradshteyn and Ryzhik 1980)
$$\Delta \phi=(b/a) F(\nu,t)\eqno (B4)$$
where $\nu = {\rm arcsin}(a/r_{min})$ and $t=d/a$. 
Substituting (B2) and (B3) in (B4) we have
$$\Delta \phi=b/a F(\pi /2 , \sqrt{2}\; {l \over b})=b/a \int -0^{\pi/2}
[1-2(l^2/b^2){\rm sin}^2\theta]^{-1/2}d \theta=\pi /2+\pi /4(l^2/b^2)$$
where we  have used the fact that $l \ll b$.
So the bending angle $\alpha $ is
$$\alpha = 2 \Delta \phi -\pi = \pi /2 (l^2/b^2).$$

\pagebreak
\begin{figure*}
\centerline{\psfig{figure=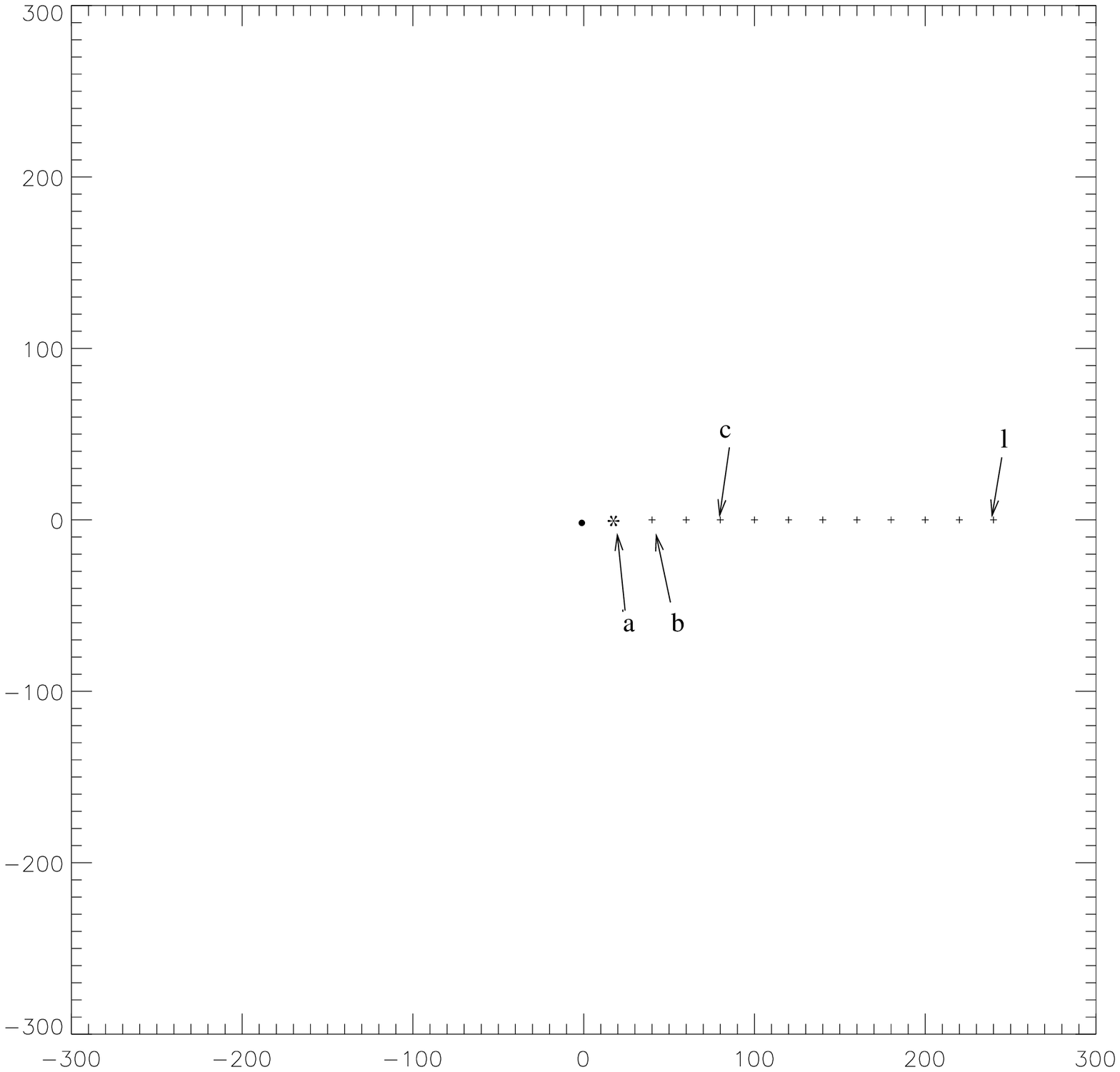,width=1.1\textwidth,angle=-180 }}
\caption{ A horizontal line of point sources in the plane of the sky .}
\end{figure*}
\begin{figure*}
\centerline{\psfig{figure=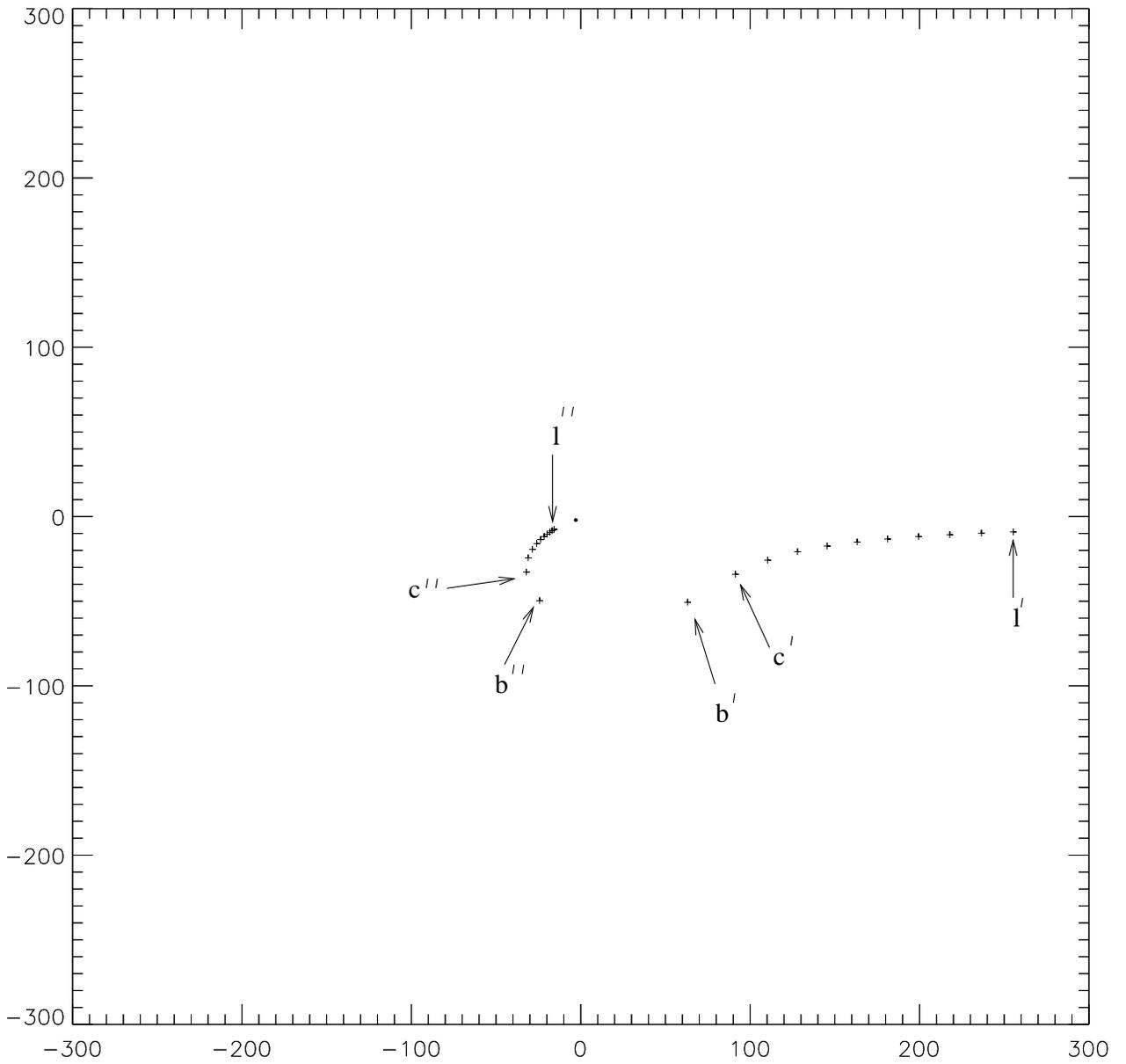,width=1.1\textwidth,angle=-90 }}
\caption{Corresponding lensed point sources for $m=Q=1$ and
 for both $\alpha <\beta$ (primed ones) and $\alpha >\beta$ (double primed
 ones) cases . Note that the point source $\it a$ can not be seen by the 
observer. Using a much denser line of point sources one will find that 
the two images form a continuous curve, i.e. there is a point  between sources 
$\it a$ and $\it b$ whose two images coincide .}
\end{figure*}
\begin{figure*}
\centerline{\psfig{figure=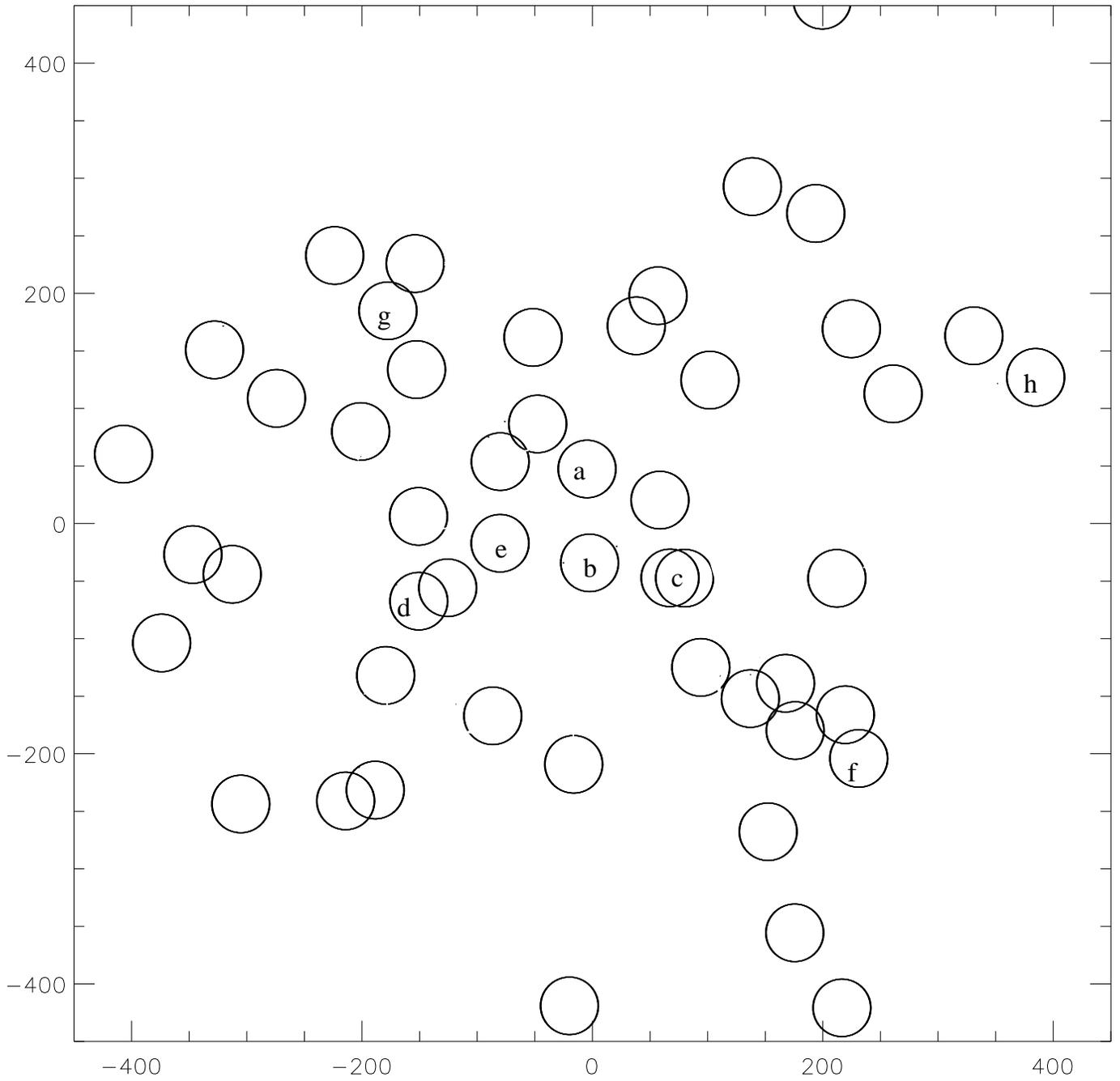,width=20cm,angle=-90 }}
\caption{ Randomly distributed circular sources in the plane of the sky .}
\end{figure*}
\begin{figure*}
\centerline{\psfig{figure=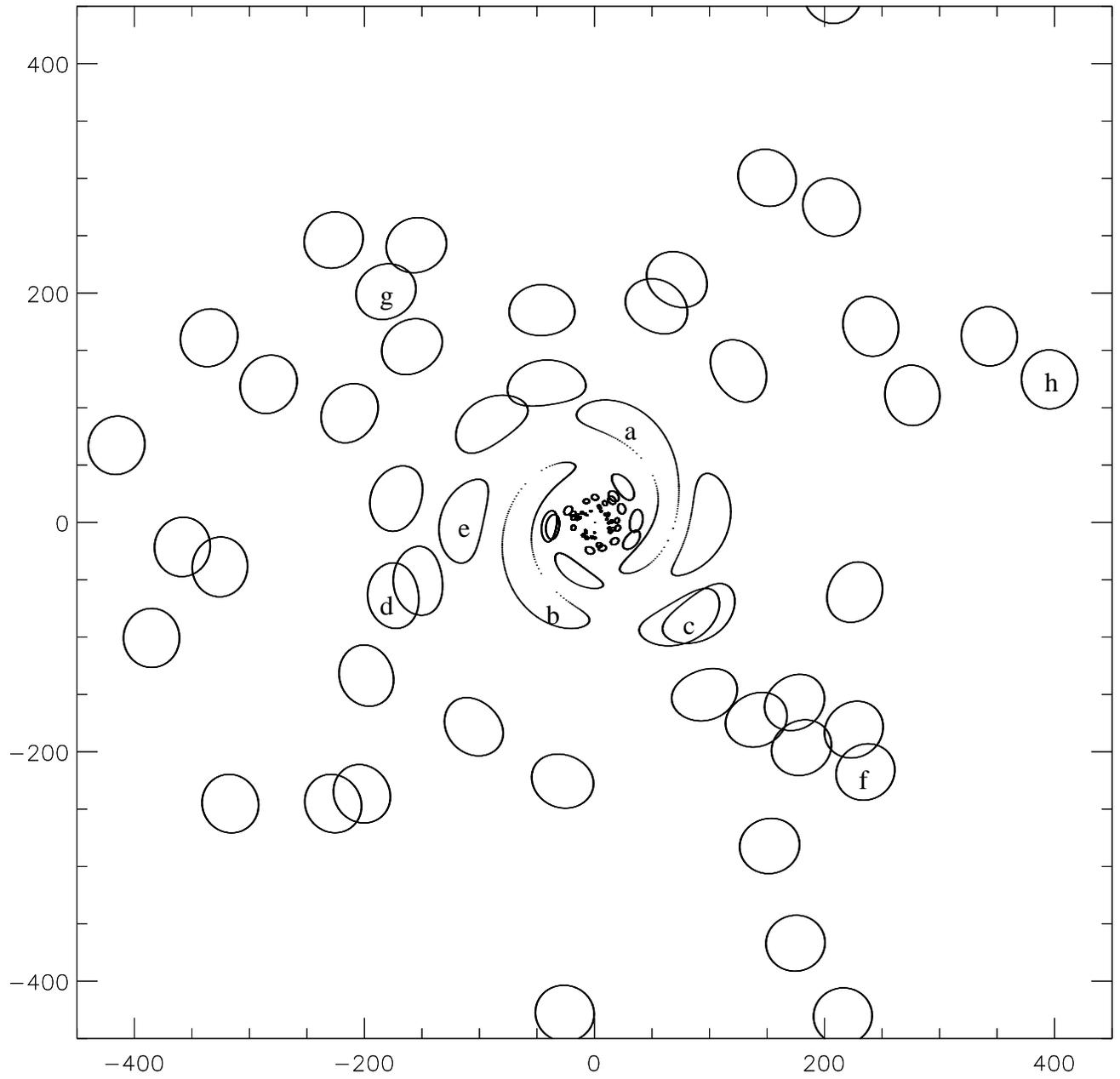,width=19cm,angle=-90 }}
\caption{  Lensed circular sources in the plane of the sky for m=Q . The
central part which is the image for the case $\alpha >\beta$ is enlarged 
in the next figure .}
\end{figure*}
\begin{figure*}
\centerline{\psfig{figure=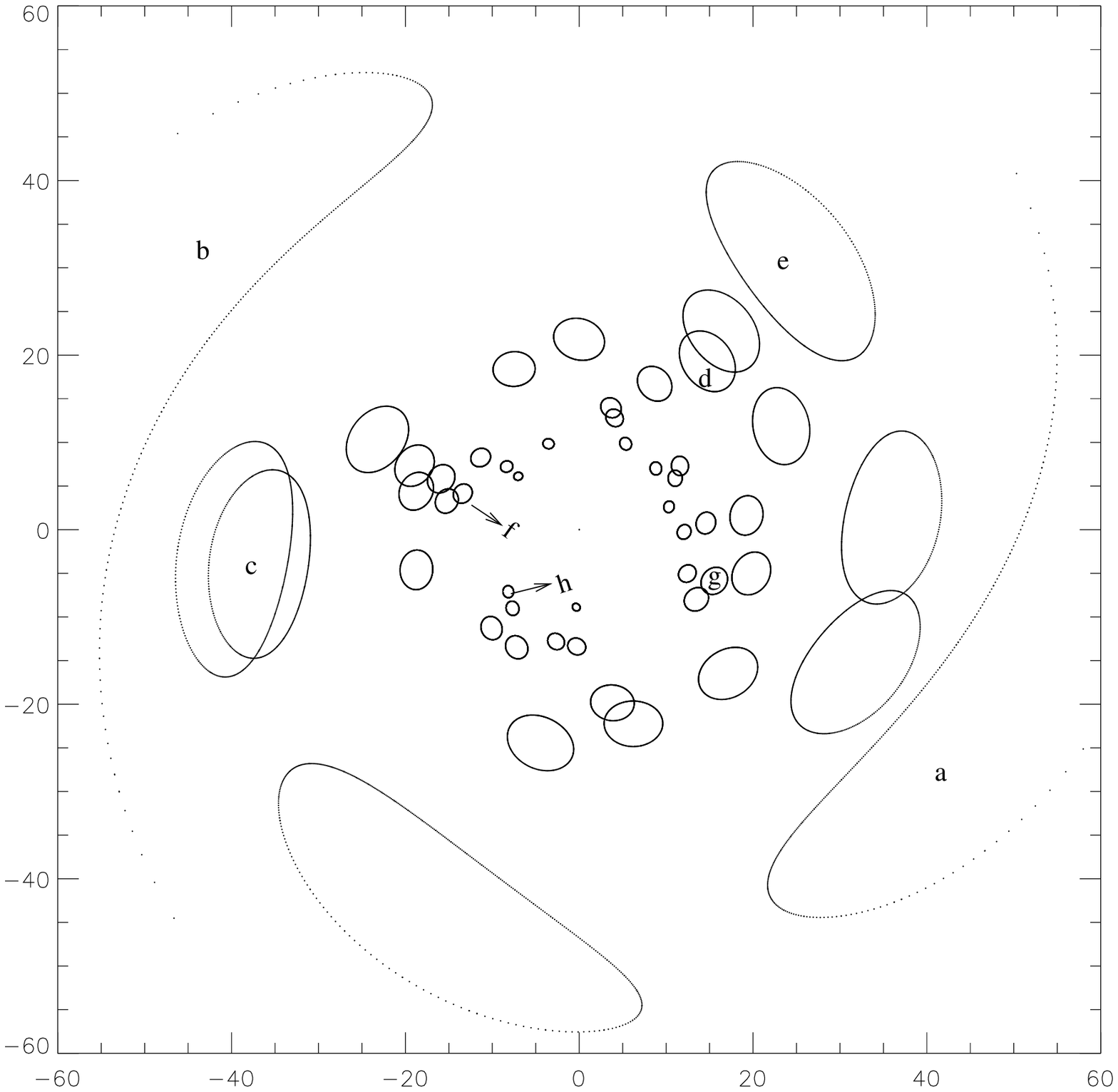,width=20cm,angle=-90 }}
\caption{  Lensed circular sources in the plane of the sky for m=Q and $\alpha
> \beta$ .  }
\end{figure*}
\begin{figure*}
\centerline{\psfig{figure=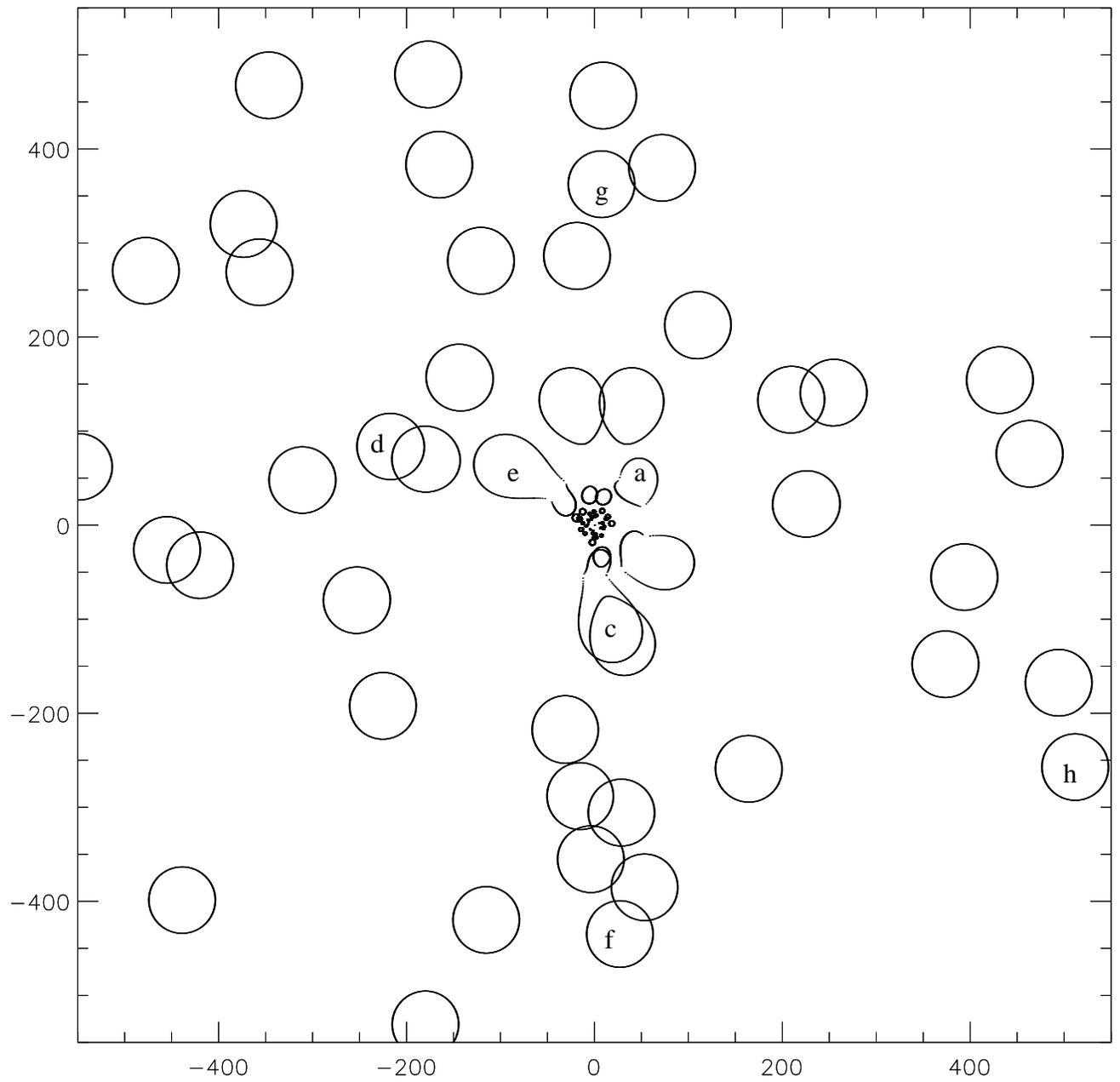,width=19cm,angle=-90 }}
\caption{  Lensed circular sources in the plane of the sky for m=0 
( i.e pure NUT ). Note that the source $b$ is completely absent in this case
 . The central part which is the image for the case 
$\alpha >\beta$ is enlarged in the next figure .  }
\end{figure*}
\begin{figure*}
\centerline{\psfig{figure=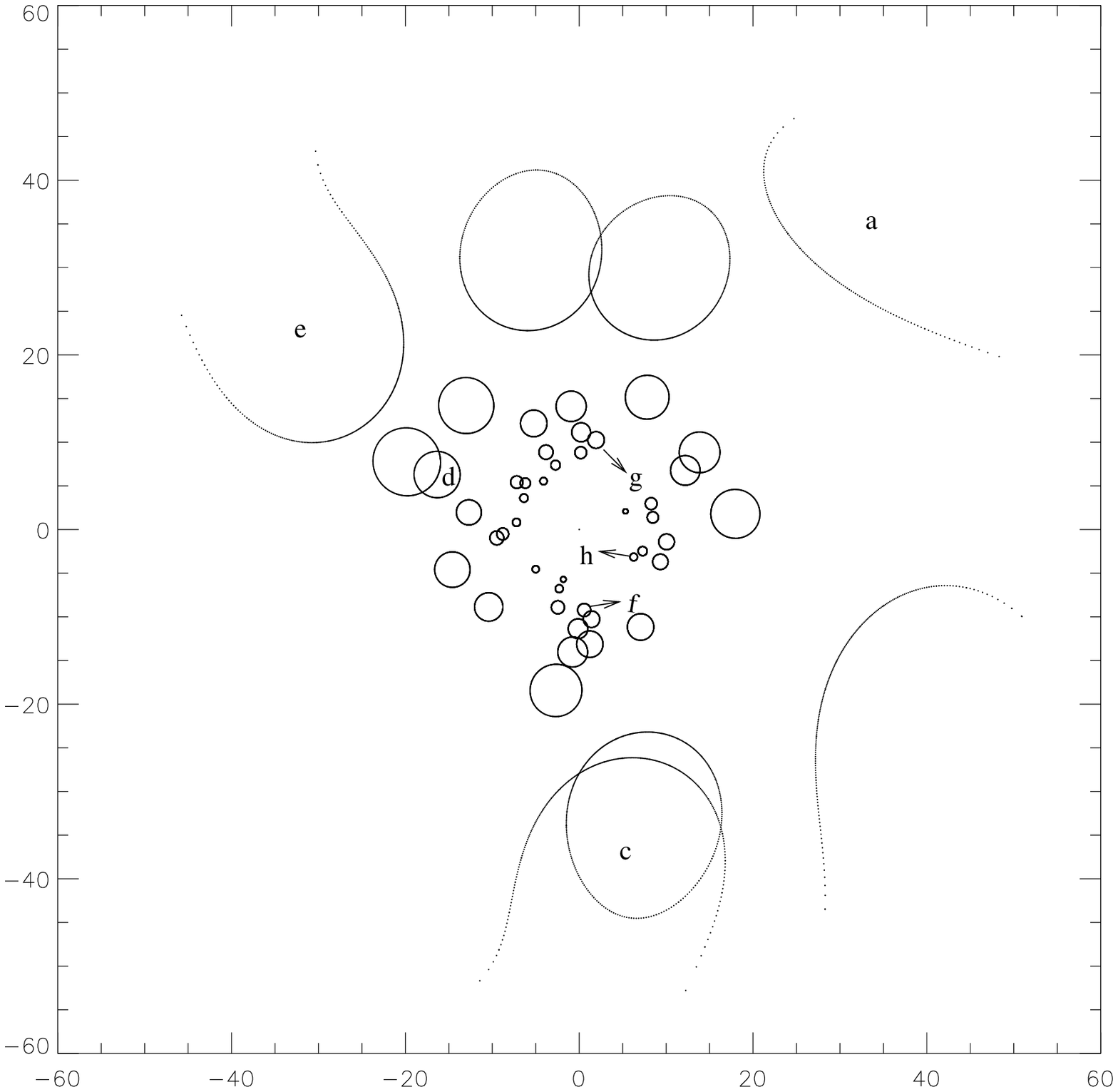,width=16cm,angle=-90 }}
\caption{  Lensed circular sources in the plane of the sky for m=0 and
 $\alpha >\beta$.  }
\end{figure*}
\medskip
\begin{figure*}
\centerline{\psfig{figure=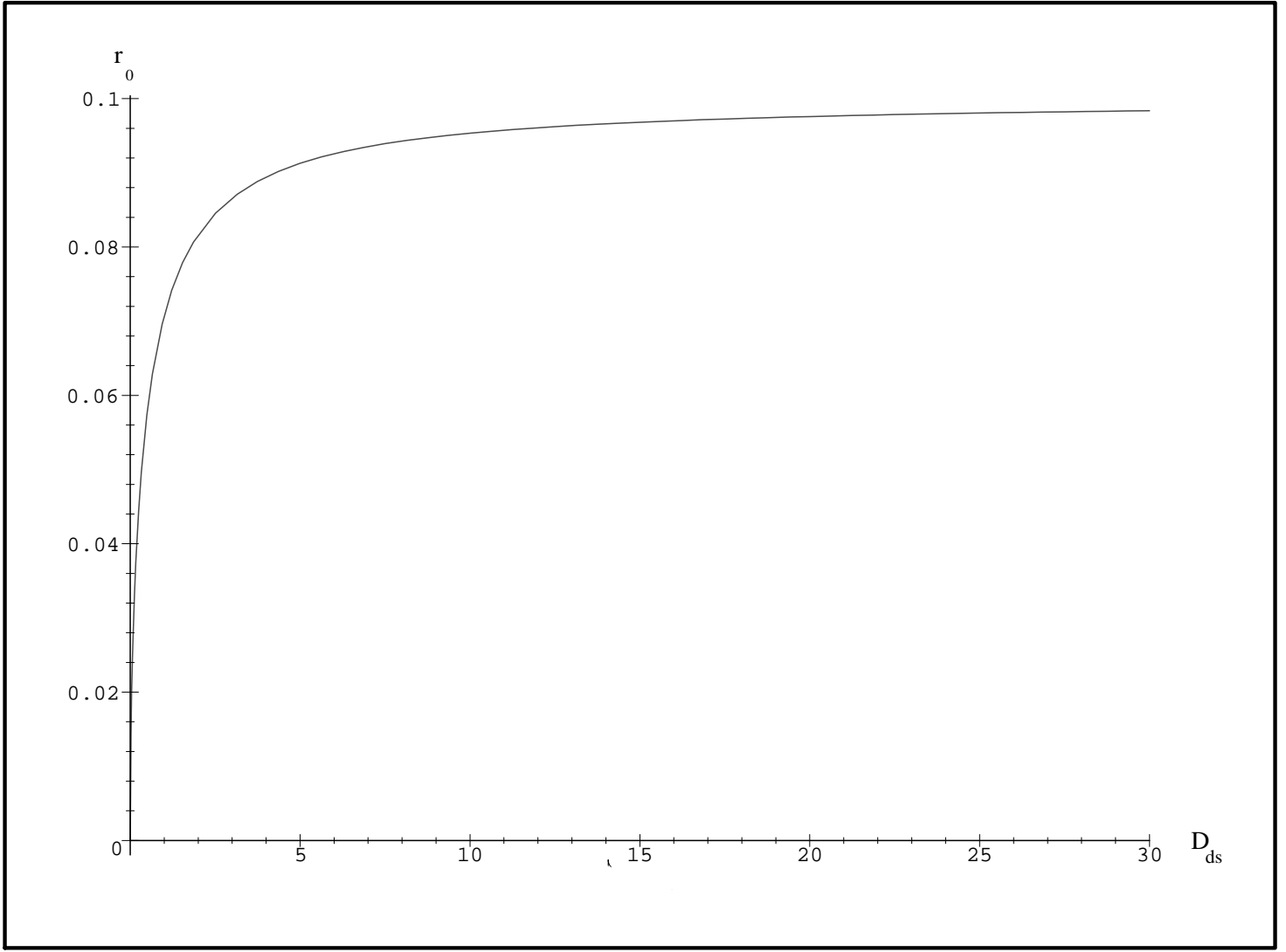,width=1.3\textwidth,angle=180 }}
\caption{$r_0$, the radius of the unseeable area in terms of the distance 
of the source to the lens, $D_{ds}$ for $m=1$, $Q=m/10$, $D_d=1$ }
\end{figure*}
\medskip
\begin{figure*}
\centerline{\psfig{figure=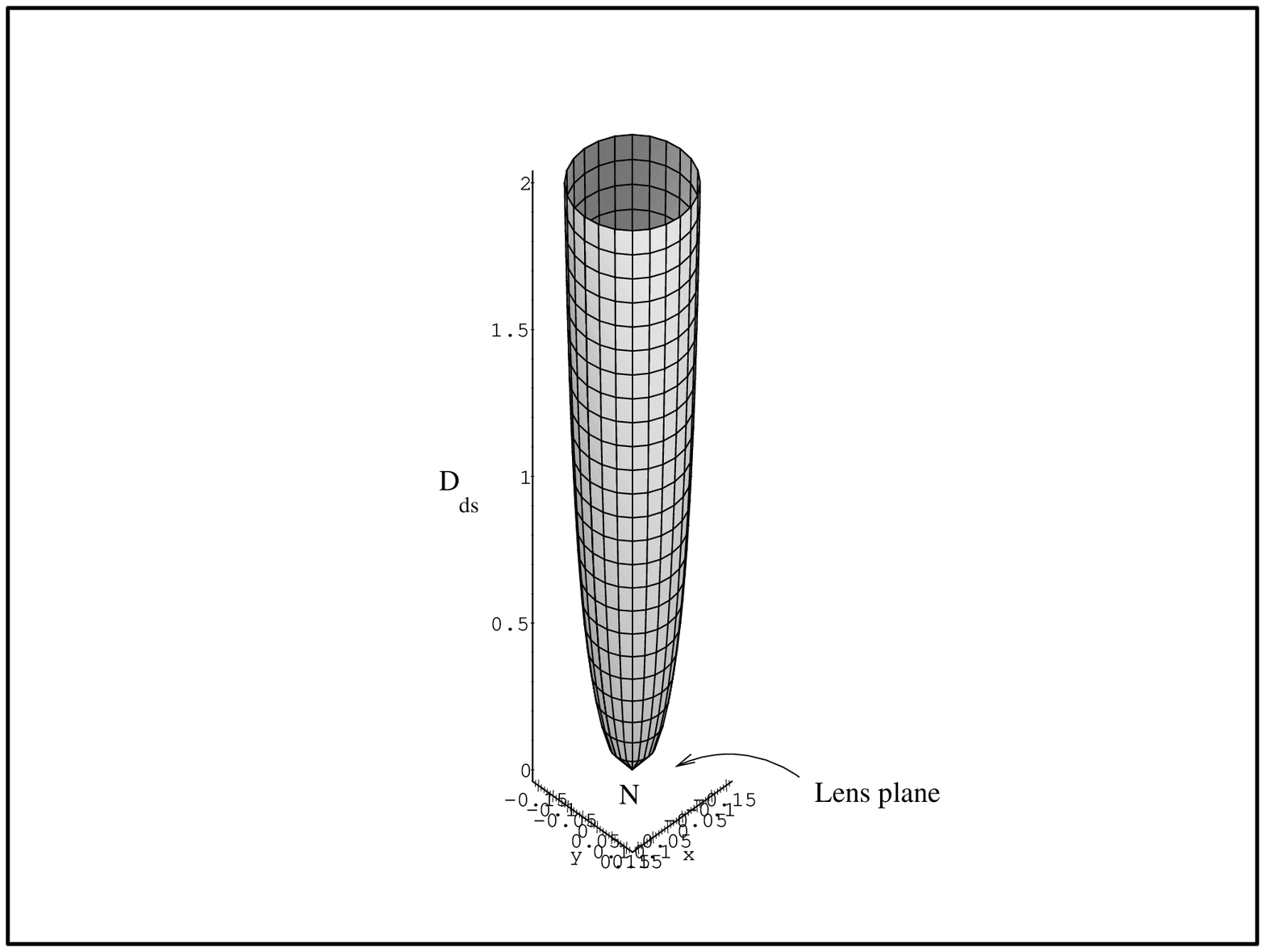,width=1.3\textwidth,angle=180 }}
\caption{ Three-dimensional version of fig. 14 for $m=1$, $Q=m/5$, $D_d=1$.
Note the difference of the scales on the Lens plane and along $D_{ds}$. }
\end{figure*}

\label{lastpage}
\end{document}